\documentclass{aa}

\usepackage[varg]{txfonts}
\usepackage[colorlinks=true, linkcolor=blue, citecolor=blue, urlcolor=gray]{hyperref}
\usepackage{xcolor}
\usepackage{listings}
\usepackage{upgreek}
\usepackage{longtable}

\bibpunct{(}{)}{;}{a}{}{,}

\newcommand{\txd}{{\text{d}}}
\newcommand{\Rb}{R_{\text{b}}}
\renewcommand{\Re}{R_{\text{e}}}
\newcommand{\Ib}{I_{\text{b}}}

\begin{document}

\title{The core of the problem: Physical limits of the core--S\'ersic model}

\titlerunning{The core--S\'ersic model}

\author{Maarten Baes}

\authorrunning{M.~Baes}

\institute{Department of Physics and Astronomy, Universiteit Gent, Proeftuinstraat 86 N3, B-9000 Ghent, Belgium, \\\email{maarten.baes@ugent.be}}

\date{Received 29 November 2025 / Accepted 5 April 2026}

\abstract{%
The core--S\'ersic model is the standard tool for describing partially depleted stellar cores in massive early-type galaxies, yet its physical admissibility has rarely been examined. Using numerical deprojections, we show that many formally allowed parameter combinations cannot represent realistic stellar systems: sharp transitions between the inner power-law core and the outer S\'ersic profile (large~$\alpha$) always generate non-monotonic intrinsic density profiles. We identify, for each set of structural parameters $(\gamma, m, \Re/\Rb)$, a critical transition parameter, $\alpha_{\text{crit}}$, above which monotonicity is violated. This threshold systematically depends on the core slope and S\'ersic index, implying that a fraction of the commonly used parameter space, including the widely adopted sharp-transition limit $\alpha\rightarrow\infty$, is physically ruled out. These constraints have important consequences for measuring core sizes and mass deficits in massive ellipticals, for constructing dynamical models, and for comparing observations with simulations of supermassive black hole binary evolution.
}

\maketitle


\section{Introduction}
 
The central regions of galaxies encode key information about their assembly histories and the co-evolution of stars and supermassive black holes (SMBHs). High-resolution imaging with the \emph{Hubble Space Telescope} (\emph{HST}) revealed that the inner surface-brightness profiles of early-type galaxies and bulges are far from universal, displaying a rich variety of cusps, shallow cores, and nuclear components \citep{Lauer1995, Faber1997, Rest2001, Laine2003, Trujillo2004b, Ferrarese2006}. Luminous elliptical galaxies and the brightest cluster galaxies typically show partially depleted stellar cores on scales of a few tens to hundreds of parsecs, where the surface brightness flattens with respect to an inward extrapolation of the outer profile. Less luminous spheroids, in contrast, tend to exhibit steep power-law cusps or compact nuclear star clusters \citep{Carollo1997, Cote2006, Ferrarese2006, Lauer2007a}. These cores and cusps are tightly intertwined with global galaxy properties and the masses of central SMBHs, and they play a central role in empirical scaling relations linking nuclear and galaxy-wide structure \citep{Graham2003, Lauer2007b, Kormendy2009, Rusli2013, Dullo2014, Dullo2019, Ferrarese2020, Quenneville2024}.

From a theoretical perspective, partially depleted cores in massive early-type galaxies are commonly interpreted as the fossil imprints of SMBH binary evolution in gas-poor mergers. After two galaxies merge, the SMBHs sink towards the common centre by dynamical friction and form a bound binary. The hardening binary ejects stars on intersecting orbits via three-body interactions, progressively scouring out a stellar core and lowering the central phase-space density \citep{Begelman1980, Milosavljevic2001, Merritt2006b, Merritt2006a, Gualandris2008}. This basic picture has been refined by increasingly sophisticated direct $N$-body and hybrid simulations that follow both the binary evolution and the response of the surrounding stellar population. These studies show that binary scouring can produce cores of the observed sizes and mass deficits, and that repeated mergers lead to progressively larger cores and distinctive kinematic signatures such as tangential anisotropy and kinematically decoupled central components \citep{Gualandris2017, Vasiliev2017, Rantala2018, Rantala2019, Rantala2024, Nasim2021, Partman2024, Khonji2024}.

On the observational side, the diversity of nuclear profiles has motivated the development of flexible parametric models to describe galaxy cores. The \citet{Lauer1995} `Nuker' model was originally introduced as a broken power law fitted to the inner regions of \emph{HST} surface-brightness profiles, and it provided a convenient phenomenological tool to classify galaxies into `core' and `power-law' types. However, the Nuker model is defined over a limited radial range, is sensitive to the chosen fitting interval, and does not connect smoothly to the global structure of a galaxy \citep{Graham2003}. The core--S\'ersic model was introduced as a more physically motivated alternative, designed to describe galaxies with partially depleted cores while retaining the successful S\'ersic description of the outer profile \citep{Graham2003, Trujillo2004b}. In this model, an inner power law of slope, $\gamma$, transitions at a break radius, $\Rb$, to an outer S\'ersic profile with index $m$; the transition sharpness is governed by a parameter, $\alpha$, and the core `strength' can be quantified by the difference between the actual profile and the inward extrapolation of the outer S\'ersic component.

The core--S\'ersic model has since become the standard tool for quantifying stellar cores in bright early-type galaxies. It is implemented in the most popular image fitting software packages, such as {\tt{GALFIT}} \citep{Peng2002, Peng2010, Bonfini2014}, {\tt{IMFIT}} \citep{Erwin2015}, and {\tt{ProFit}} \citep{Robotham2017}. It has been used to measure core sizes and mass deficits in large samples and to calibrate scaling relations between core properties and SMBH masses \citep{Ferrarese2006, Ferrarese2020, Hyde2008, Richings2011, Rusli2013, Dullo2013, Dullo2014, Bonfini2015, Bonfini2016, Dullo2017, Dullo2018, Dullo2023, denBrok2021}. Core--S\'ersic fits underpin many current constraints on the efficiency of binary scouring, the cumulative number of major dry mergers experienced by massive ellipticals, and the role of gravitational-wave recoil in expanding or refilling cores. In this sense, the core--S\'ersic model plays a central role in connecting observed galaxy cores to the formation and dynamical evolution of SMBH binaries.

Despite its widespread use, the core--S\'ersic model is usually treated purely as a fitting function for surface-brightness profiles. For applications that require intrinsic (three-dimensional) densities or dynamical models, however, the full deprojected profile is needed. Even at the level of the density profile, not all formal combinations of core--S\'ersic parameters necessarily correspond to realistic stellar systems. For example, some parameter choices may lead to unphysical behaviour of the intrinsic density at small or large radii, to density profiles that are not strictly monotonic, or to models with divergent total mass. These limitations are especially relevant when core--S\'ersic parameters are used in dynamical modelling, in comparisons with $N$-body or hydrodynamical simulations, or as priors in Bayesian fitting codes.

In this paper we therefore revisit the core--S\'ersic model from the point of view of intrinsic density profiles and, more generally, of basic physical plausibility. Building on our previous work on the deprojection of the S\'ersic models \citep{Baes2011a, Baes2011c, Baes2019b, Baes2019d} and Nuker models \citep{Baes2020a}, we investigate which regions of the core--S\'ersic parameter space correspond to stellar systems with realistic three-dimensional density distributions. Our goals are threefold. First, we derive and discuss the intrinsic density profiles associated with the general core--S\'ersic family, clarifying their behaviour at small and large radii. Second, we identify combinations of parameters that lead to pathologies in the density profile (such as non-monotonic behaviour) and delineate the subspace of parameters that yield physically acceptable models. Third, we provide practical guidance for the use of core--S\'ersic fits in observational and theoretical studies of galaxy cores, in particular by proposing a set of constraints and priors on the model parameters that ensure realistic intrinsic density profiles. In this way, the widely used core--S\'ersic model can be placed on a firmer physical footing and used more reliably as a bridge between observed surface-brightness profiles, stellar dynamics, and galaxy formation models.

\begin{figure*}
\includegraphics[width=0.96\textwidth]{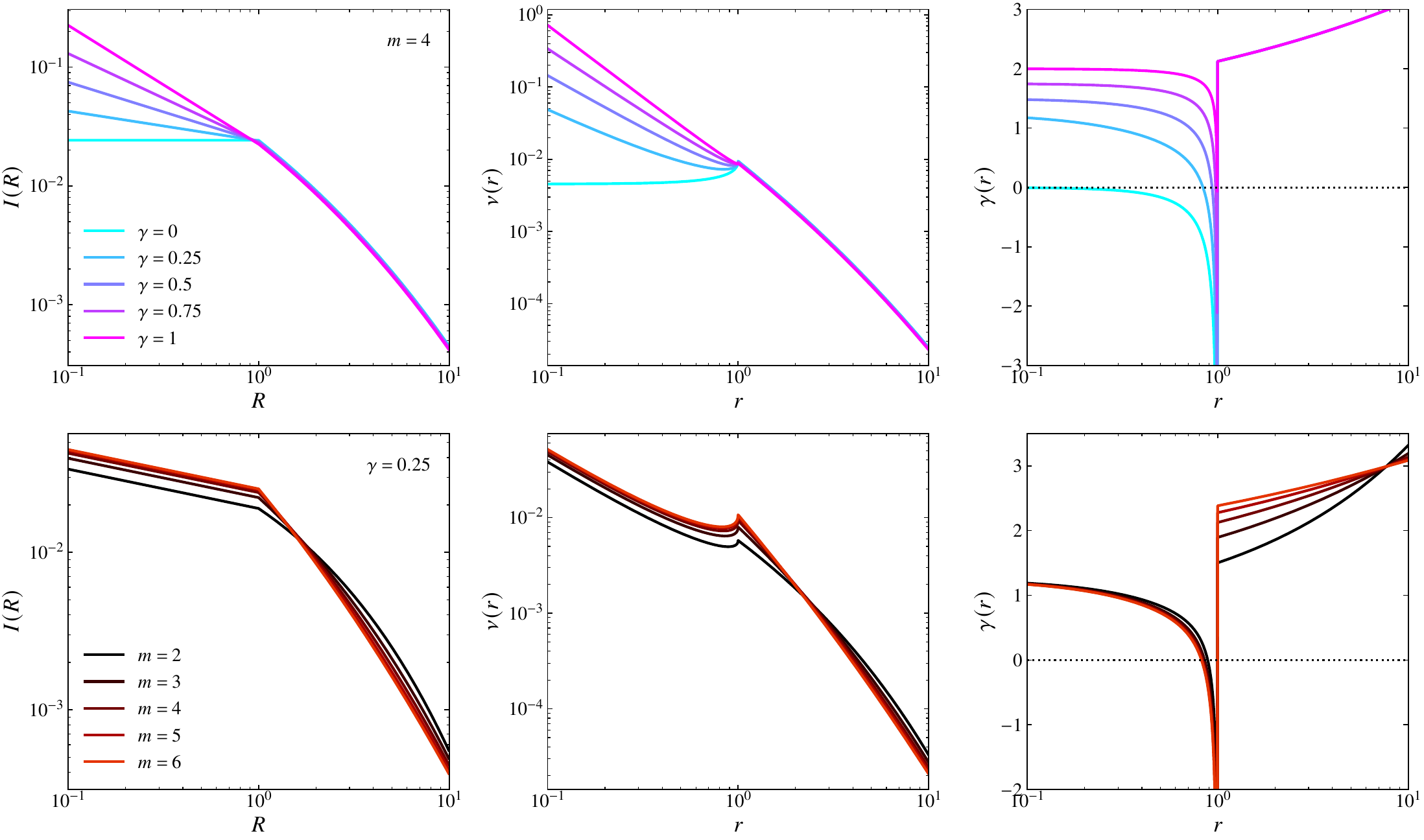}%
\caption{Surface-brightness profiles (left), intrinsic luminosity density profiles (middle), and logarithmic density slopes (right) for sharp-transition core--S\'ersic models ($\alpha \rightarrow \infty$). All models have a total luminosity of $L = 1$, a break radius of $\Rb = 1$, and an effective radius of $\Re = 5$. In the top row the S\'ersic index is fixed to $m = 4$, while the inner power-law slope, $\gamma$, is varied as indicated; in the bottom row the inner slope is fixed to $\gamma = 0.25$ and the S\'ersic index $m$ is varied. The projected profiles show the expected behaviour of a pure power law for $R \leq \Rb$ joined to a S\'ersic profile for $R \geq \Rb$, with a discontinuity in the radial derivative at $\Rb$. This discontinuity translates into a sharp peak in the intrinsic density at $r \approx \Rb$ and a region with $\gamma(r) < 0$ just inside the break radius, demonstrating that all sharp-transition core--S\'ersic models have non-monotonic and hence unphysical intrinsic density profiles.}
\label{SharpCoreSersic.fig}
\end{figure*}

\section{The core--S\'ersic model}

The core--S\'ersic model is defined by the following surface brightness profile:
\begin{subequations}
\label{coreSersic}
\begin{equation}
I(R) 
= 
I'
\left[1+\left(\frac{R}{\Rb}\right)^{-\alpha}\right]^{\frac{\gamma}{\alpha}}
\exp\left[-b\left(\frac{R^\alpha+\Rb^\alpha}{\Re^\alpha}\right)^{\frac{1}{\alpha m}}\right]
,\end{equation}
with
\begin{equation}
I' = \Ib\,2^{-\frac{\gamma}{\alpha}} 
\exp\left[b\left(\frac{2\Rb^\alpha}{\Re^\alpha}\right)^{\frac{1}{\alpha m}}\right].
\end{equation}
\end{subequations}
The parameter $\Rb$ in this formula represents the break radius and separates the inner and outer profiles. At small radii, $R\ll\Rb$, the core--S\'ersic profile behaves as a power law with a negative logarithmic slope, $\gamma$. At large radii, $R\gg\Rb$, the model behaves as a S\'ersic profile with half-light radius, $\Re$, and S\'ersic index, $m$. The two profiles join smoothly at $R=\Rb$, and $\Ib=I(\Rb)$ represents the intensity at the break radius. The last parameter, the dimensionless transition parameter, $\alpha$, sets the sharpness of the transition between the inner power-law behaviour and the outer S\'ersic-like behaviour, in a similar way as in the Nuker model. In general, the core--S\'ersic model is thus fully characterised by the six parameters $(\Ib, \Rb, \gamma, \Re, m, \alpha)$, or equivalently $(L, \Rb, \gamma, \Re, m, \alpha)$, with $L$ the total luminosity. The quantity $b \equiv b(m)$ in Eq.~(\ref{coreSersic}) is not a free parameter, but a dimensionless number that is inherited from the S\'ersic model, and that guarantees that $\Re$ is the effective radius.\footnote{Note that we followed \citet{Bonfini2014} in choosing $\Re$ as the effective radius of the S\'ersic part of the profile, and not the effective radius of the entire core--S\'ersic model. In other words, $\Re$ is not the radius of the isophote that contains half of the total luminosity of the intensity profile~(\ref{coreSersic}). It is possible to redefine $\Re$ as the effective radius of the core--S\'ersic model by rescaling $b$ \citep{Trujillo2004b}.} Suitable approximations and numerical values for $b(m)$ can be found in \citet{Ciotti1999} and \citet{Baes2019b}. 

The sharp-transition core--S\'ersic model corresponds to $\alpha\to\infty$, which simplifies the surface brightness profile to 
\begin{equation}
I(R) 
= 
\begin{cases}
\;\displaystyle
\Ib \left(\frac{R}{\Rb}\right)^{-\gamma}
&
\quad R\leqslant\Rb,
\\[1em]
\;\displaystyle
\Ib \exp\left[-b\left(\frac{R}{\Re}\right)^{\frac{1}{m}} + b\left(\frac{\Rb}{\Re}\right)^{\frac{1}{m}}\right]
&
\quad R\geqslant\Rb.
\end{cases}
\end{equation}
In this case, we have a pure power-law profile at $R\leqslant\Rb$ and a pure S\'ersic profile at $R\geqslant\Rb$, which join at the break radius. The model is completely defined by the set of five parameters $(\Ib, \Rb, \gamma, \Re, m)$, or equivalently $(L, \Rb, \gamma, \Re, m)$.

\section{Luminosity density}

\begin{figure*}
\includegraphics[width=0.96\textwidth]{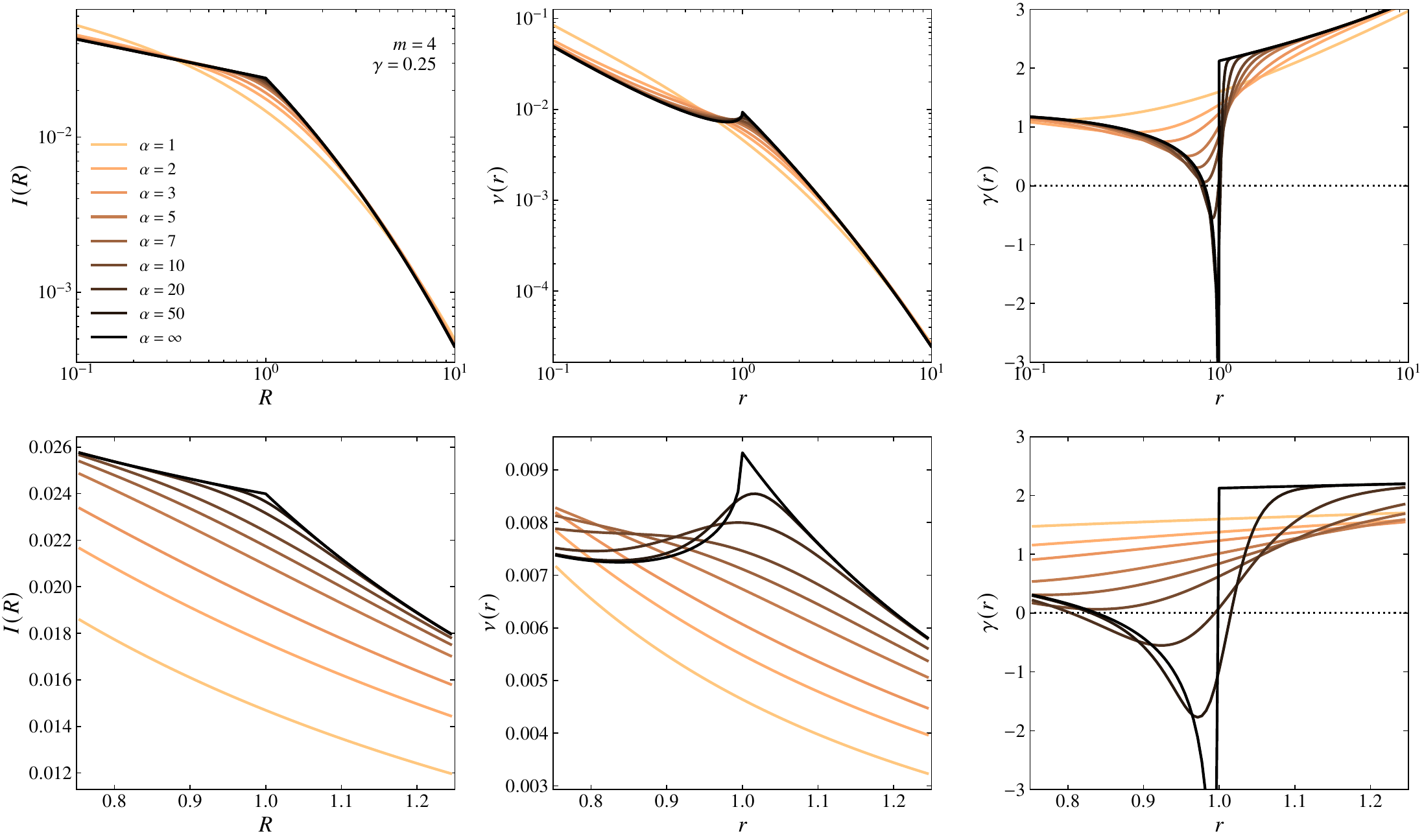}%
\caption{Intrinsic luminosity density profiles (left) and logarithmic density slopes $\gamma(r)$ (right) for a sequence of core--S\'ersic models with increasingly smooth transitions. All models share the same structural parameters as in Fig.~1, but the transition parameter $\alpha$ takes the values $\alpha = 4, 2, 1, 0.5, 0.25,$ and $0.1$, as indicated. Large values of $\alpha$ produce a strong peak in the intrinsic density just outside the break radius, reflecting the discontinuity in the derivative of the projected profile. As the transition becomes smoother (smaller $\alpha$), this peak gradually diminishes and shifts outwards, and the region with negative logarithmic slope $\gamma(r)$ shrinks. For sufficiently small $\alpha$, the density profile becomes fully monotonic and physically admissible. The comparison illustrates that the transition parameter, rather than the inner or outer slopes themselves, primarily determines whether a core--S\'ersic model yields a realistic three-dimensional stellar density distribution.}
\label{CoreSersic.fig}
\end{figure*}

\subsection{Calculation of the luminosity density profile}

Assuming spherical symmetry, the luminosity density, $\nu(r)$, of a model with a surface brightness profile, $I(R)$, was calculated through an inverse Abel transform,
\begin{equation}
\nu(r) = -\frac{1}{\pi} \int_r^\infty \frac{\txd I(R)}{\txd R}\,\frac{\txd R}{\sqrt{R^2-r^2}}.
\end{equation}
For both the Nuker and S\'ersic models, this integral cannot be evaluated in terms of elementary functions or standard special functions. A formal expression can be found in terms of the Fox $H$ function \citep{Mazure2002, Baes2011a, Baes2011c, Baes2020a}. For the core--S\'ersic model, an analytical evaluation of this integral is impossible.

To calculate the luminosity density of the core--S\'ersic model, we made use of {\tt{SpheCow}} \citep{Baes2021b}, a C++ package designed to investigate the dynamical structure of galaxies and dark matter haloes. The code uses an efficient high-order Gauss-Legendre numerical integration strategy. {\tt{SpheCow}} can generate dynamical models based on any density or surface density profile, and the user can choose between isotropic, radial, or Osipkov--Merritt orbital structures. 

We implemented two new classes, {\tt{CoreSersicModel}} and {\tt{SharpCoreSersicModel}}, in the {\tt{SpheCow}} package. For models based on their surface density profile, such as these, the only functions that need to be implemented are the surface density profile and its first, second, and third derivatives. This allows for the full calculation of the dynamical structure \citep[see Sect.~2.2 of][]{Baes2021b}. For the present work, we focus on the calculation of the luminosity density profile rather than on investigating the full dynamical structure.

\subsection{Sharp-transition core--S\'ersic model}

In Fig.~{\ref{SharpCoreSersic.fig}} we show the deprojection of two sets of sharp-transition core--S\'ersic models. All models shown have $L=1$, $\Rb=1$, and $\Re = 5$. In the top row we fixed $m=4$ and varied the power-law index, $\gamma$; in the bottom row we fixed $\gamma=0.25$ and we varied the S\'ersic index, $m$. On each row, the left panel shows the surface brightness profile, which shows the characteristic behaviour of a pure power law for $R\leq\Rb$ and a S\'ersic model for $R\geq\Rb$. Both profiles meet at $R=\Rb$. The derivative of the surface brightness profile is discontinuous at this radius. This discontinuity translates into the particular behaviour in the luminosity density shown in the panels on the middle column. Rather than a smoothly decreasing behaviour, the luminosity density has a sharp peak at $R=\Rb$. This peculiar behaviour is demonstrated in a different way in the panels on the right column, which show the (negative) luminosity density slope,
\begin{equation}
\upgamma(r) = -\frac{\txd \log \nu(r)}{\txd \log r}.
\end{equation}
All sharp-transition core--S\'ersic models have $\upgamma(r) < 0$ for $r \lesssim \Rb$, meaning that the luminosity density increases with increasing radius just before the break radius. 

\subsection{Soft-transition core--S\'ersic model}

On the top row of Fig.~{\ref{CoreSersic.fig}} we show similar plots, but now for the more general core--S\'ersic family with a smooth transition. The models shown all have fixed parameters $L=1$, $\Rb=1$, $\Re = 5$, $\gamma=0.25$, and $m=4$, and only differ in the value of the transition parameter, $\alpha$. The top left panel clearly shows the effect of this parameter on the surface brightness profile: for small values of $\alpha$, the inner and outer profile are smoothly connected, whereas this transition becomes sharper when $\alpha$ increases. In the limit $\alpha\to\infty$, the model reduces to the sharp-transition core--S\'ersic model. 

The top central and top right panels show the corresponding luminosity density profiles. For small $\alpha$, the luminosity density smoothly decreases over the entire radial range, resulting in a positive $\upgamma(r)$. As $\alpha$ increases, the luminosity density becomes increasingly flat at radii of $r\lesssim\Rb$, resulting in increasingly small values of the density slope around the break radius. This is more clearly visible in the panels of the bottom row, which show the same models and the same quantities, but now in linear scaling and zoomed in on a small range around the break radius. If $\alpha$ is sufficiently large -- that is, if the transition is sufficiently sharp -- the luminosity profile becomes non-monotonic\footnote{The occurrence of non-monotonicity is not a numerical artefact, but arises from the structure of Eq.~(\ref{coreSersic}) itself. We have checked our results using numerical integration with Mathematica.} and $\upgamma(r)<0$ for $r\lesssim\Rb$. For the set of parameters chosen in this figure, this can clearly be noted for the models with $\alpha=20$ and $\alpha=50$. Finally, in the limit $\alpha\to\infty$, we obtain a luminosity profile that shows a sharp peak at $r=\Rb$, as has already been observed in Fig.~{\ref{SharpCoreSersic.fig}}.

\begin{figure*}
\sidecaption
\includegraphics[width=12.5cm]{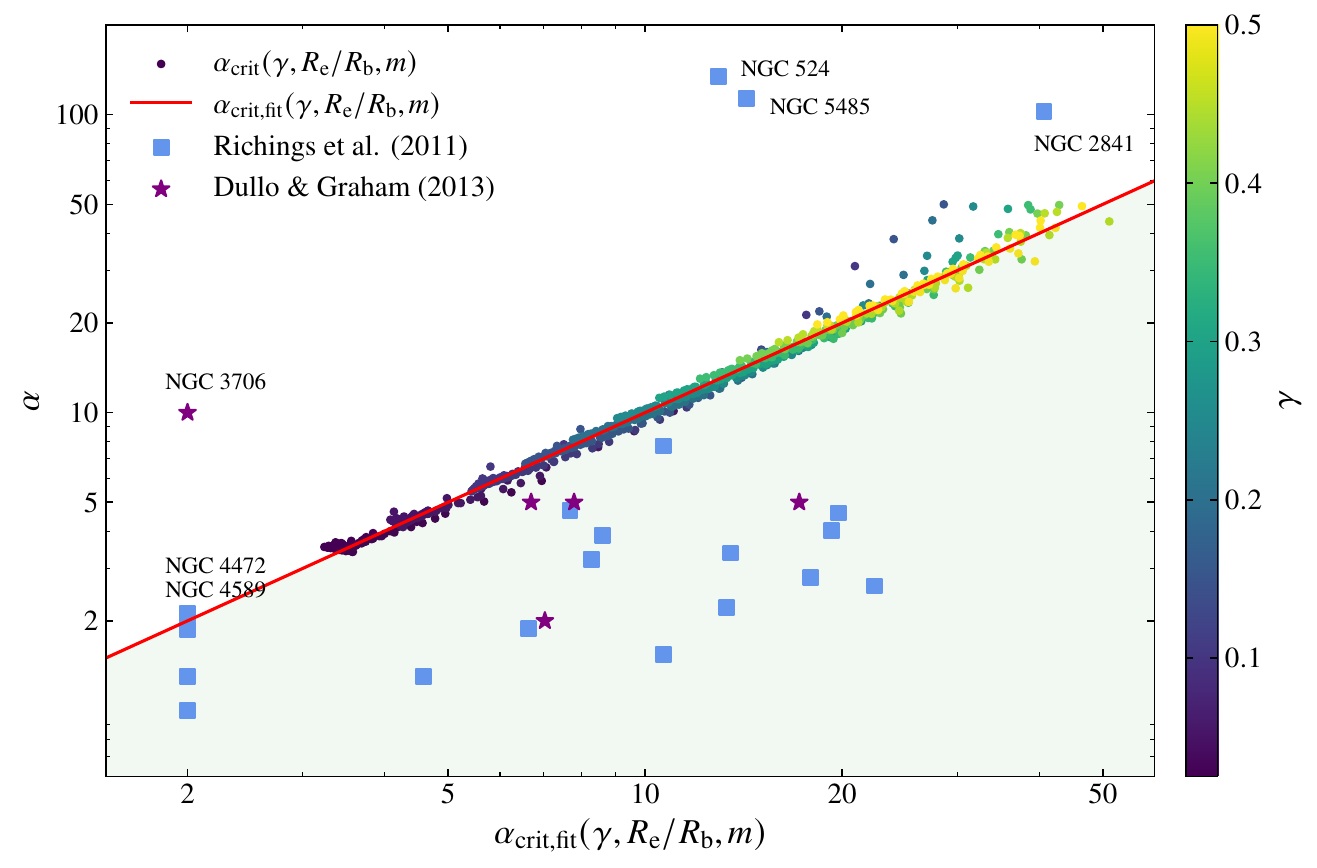}
\caption{Comparison between the transition parameter, $\alpha$, and the critical value, $\alpha_{\text{crit}}(\gamma, \Re/\Rb, m)$, that separates physically admissible and inadmissible core--Sérsic models. The horizontal axis shows the fitted approximation $\alpha_{\text{crit,fit}}(\gamma, \Re/\Rb, m)$, while the vertical axis shows the actual transition parameter, $\alpha$. The red line indicates the boundary $\alpha = \alpha_{\text{crit}}$: models below this line have monotonically decreasing intrinsic density profiles and are therefore physically admissible, whereas models above the line produce non-monotonic densities and are unphysical. The colour scale indicates the inner logarithmic slope, $\gamma$. Overplotted are core--S\'ersic parameters from the observational studies of \citet[][blue squares]{Richings2011} and \citet[][violet stars]{Dullo2013}, illustrating that most observed systems lie within the physically allowed region, while a few fall in the inadmissible part of the parameter space.}
\label{alphacrit.fig}
\end{figure*}

In summary, for every set of parameters $(\gamma, m, \Re/\Rb)$, there is a critical transition parameter, $\alpha_{\text{crit}}$, that forms the boundary between models with monotonic and non-monotonic density profiles. We used {\tt{SpheCow}} to calculate this critical transition parameter for a large suite of core--S\'ersic models with different values for $\gamma$, $m$, and $\Re/\Rb$. The results are listed in Table~{\ref{tab:alphacrit}} in Appendix~{\ref{appA.sec}}. For $\gamma=0$, we always find $\alpha_{\text{crit}} = 2$; as $\gamma$ increases, $\alpha_{\text{crit}}$ increases too. At fixed $\gamma$, $\alpha_{\text{crit}}$ is a decreasing function of $m$ and an increasing function of $\Re/\Rb$. We approximated $\alpha_{\text{crit}}(\gamma, \Re/\Rb, m)$ by means of an analytical function, 
\begin{multline}
\label{fit}
\ln(\alpha_{\text{crit,fit}} - 2) 
= 
\left(a_0 + \frac{a_1}{m} + a_2\ln\gamma\right) 
+
\left(a_3 + \frac{a_4}{m} + a_5\ln\gamma\right) \gamma \\
+ 
\left(a_6 + \frac{a_7}{m} + a_8\ln\gamma\right) \ln\left(\frac{\Re}{\Rb}\right), 
\end{multline}
with fitting parameters
\begin{align}
\nonumber 
\quad a_0 &=  2.0188, &
\quad a_1 &= -2.6199, &
\quad a_2 &=  0.3299, \quad \\
\quad a_3 &=  1.2724, &
\quad a_4 &=  8.2291, &
\quad a_5 &=  0.0576, \\
\quad a_6 &=  0.0722, &
\quad a_7 &=  1.2773, &
\quad a_8 &=  0.0763.
\nonumber
\end{align}
This fitting function has a root-mean-square fractional error of 5.96\% over the entire range of parameters investigated. For practical applications, Eq.~(\ref{fit}) provides a convenient approximation of the critical transition parameter and can be used to impose physically motivated priors in fitting procedures. Figure~{\ref{alphacrit.fig}} compares the numerically determined values for $\alpha_{\text{crit}}(\gamma, \Re/\Rb, m)$ with the analytical fit (\ref{fit}) for all the models listed in Table~{\ref{tab:alphacrit}}. The solid red line is the one-to-one line. Models below this red line, i.e. in the green area of the plot, satisfy the relation $\alpha \le \alpha_{\text{crit}}$, and hence have a monotonically decreasing density profile. Models above the red line have $\alpha > \alpha_{\text{crit}}$ and can therefore be considered unphysical.

Figure~{\ref{alphacrit.fig}} also contains data from two representative observational studies.  \citet{Richings2011} considered a set of 150 active galaxies from the Palomar spectroscopic survey with available {\it{HST}} imaging data and fitted S\'ersic, core--S\'ersic, and double S\'ersic models to the surface brightness profiles. The surface brightness fits were considered confident for 62 galaxies, and for 21 of them, the core--S\'ersic model provided the best fit. These 21 galaxies are shown as blue squares in Fig.~{\ref{alphacrit.fig}}. The value of $\alpha$ was allowed to vary freely in the fitting procedure, leading to widely varying values, ranging from 1 to 134. \citet{Dullo2013} considered a small sample of six lenticular galaxy candidates, and fitted a combined core--S\'ersic plus exponential disc model to the surface brightness profiles for five of them. They considered three discrete values for $\alpha$: 10, 5, and 2, representing sharp, moderate, and broad transition regions, respectively \citep{Dullo2012}. These five galaxies are shown as the violet stars in Fig.~{\ref{alphacrit.fig}}.

The majority of the core galaxies from \citet{Richings2011} and \citet{Dullo2013} are in the physically admissible part of the parameter space. Six galaxies, however, labelled with their NGC numbers, are located above the red line, and thus correspond to non-monotonic density profiles.

\section{Discussion}
\label{sec:discussion}

The structure of partially depleted stellar cores in massive early-type galaxies encodes the cumulative impact of galaxy mergers and the dynamical evolution of SMBH binaries. In this context, the core--S\'ersic model has become the standard tool for quantifying the inner slope, break radius, and luminosity deficit of such systems. Our results show, however, that the commonly adopted parameterisation of the model does not always correspond to a physically meaningful three-dimensional stellar system. The constraints derived in this work therefore have direct implications for both the interpretation of core structure and the practical use of core--S\'ersic fits in observational studies.

A central theme that emerges from our analysis is that the transition parameter, $\alpha$, plays a far more restrictive role than is often appreciated. While the intuitive expectation is that $\alpha$ merely modulates how smoothly the inner and outer components join, we find that for a broad range of S\'ersic indices, $m$, inner slopes, $\gamma$, and break radii, $\Rb$, sharp transitions inevitably produce non-monotonic intrinsic density profiles. This behaviour is unphysical for any realistic collisionless stellar system, where the density is expected to decrease monotonically with radius outside the very nucleus. In particular, the sharp-transition limit $\alpha \rightarrow \infty$, frequently adopted in the literature to simplify the fitting procedure or reduce parameter degeneracies, is never physically admissible. This result parallels, but is not identical to, the conclusion reached previously for the Nuker model \citep{Baes2020a}, reinforcing that broken power-law transitions, even when embedded in a global S\'ersic structure, must remain sufficiently gradual to preserve realistic intrinsic behaviour.

Our analysis assumes spherical symmetry when deprojecting the surface-brightness profile to obtain the intrinsic luminosity density. Real galaxies may of course be flattened and observed under an arbitrary inclination, which introduces additional degeneracies in the deprojection from projected to intrinsic structure. However, the non-monotonic behaviour identified here originates from the sharp transition in the projected core--S\'ersic profile itself rather than from the assumed geometry. Changing the inclination or adopting a mildly axisymmetric intrinsic shape would modify the quantitative details of the recovered density profile but would not remove the characteristic peak near the break radius produced by sufficiently large transition parameters. If a given set of core--S\'ersic parameters yields a non-monotonic density profile under the simplest spherical deprojection, it cannot correspond to a physically plausible stellar distribution under more general geometries either.

It is important to emphasise that monotonicity of the intrinsic density profile plays an important role in assessing the consistency of dynamical models. Full phase-space consistency requires the distribution function to remain non-negative over the entire phase space. For models with an isotropic orbital structure, monotonicity of the density profile is a necessary, though not by itself sufficient, condition for consistency \citep{Binney2008, Ciotti2021}. For models with a more general anisotropic orbital structure, identifying necessary conditions for the phase-space consistency is notoriously difficult \citep[e.g.][]{An2006, An2012, Baes2007}. For models with Osipkov--Merritt or Cuddeford orbital structures, among the most widely used orbital structures beyond the isotropic case, monotonicity of the density profile is also a necessary condition \citep{Ciotti1992, Ciotti2010}. Our focus on monotonicity therefore targets the most immediate and robust indicator of physical admissibility for the core--S\'ersic family.

The critical transition parameter, $\alpha_{\text{crit}}$, that we identify for each $(\gamma, m, \Re/\Rb)$ combination delineates the physically allowed region of the core--S\'ersic parameter space. The general trend is that steeper outer profiles (larger $m$) and flatter inner slopes (smaller $\gamma$) require particularly smooth transitions, because they otherwise force the density to decrease too quickly outside the break radius before turning upwards again as the S\'ersic component dominates. Conversely, for shallower components the allowed parameter space widens but remains bounded: even moderate transition parameters can violate monotonicity for typical structural parameters of luminous core ellipticals. These systematic trends highlight that the degeneracy between $\alpha$ and the other core--S\'ersic parameters is not merely a nuisance in fitting, but a fundamental constraint on the existence of an underlying physical stellar distribution.

These findings directly influence the interpretation of core sizes and mass deficits, two quantities widely used to test predictions of SMBH binary scouring models. Because the inferred break radius and luminosity deficit depend sensitively on the adopted transition parameter, fits that implicitly or explicitly assume large $\alpha$ may systematically overestimate the abruptness of the profile change, thereby biasing estimates of the evacuated stellar mass. In particular, our results suggest that the sharp-transition fits sometimes used in the literature may yield mass deficits that are inconsistent with any realistic stellar density distribution, independent of the quality or radial extent of the observational data. Such biases propagate into derived relations between core size, galaxy mass, and SMBH mass, and can affect comparisons with both idealised $N$-body simulations and cosmological galaxy-formation models \citep[e.g.][]{Merritt2006a, Gualandris2008, Rantala2018, Rantala2019, Gualandris2017, Vasiliev2017}.

More broadly, our work emphasises that the core--S\'ersic model must be treated as a physically interpretable three-dimensional model, not merely a flexible fitting function for the surface brightness distribution. The distinction is crucial when the model is used as input for dynamical modelling, for constructing initial conditions in simulations, or when seeking to relate observed core properties to the merger-driven evolutionary pathways predicted by theory. The strict constraints we derive provide guidance for such applications: fits obtained with transition parameters exceeding $\alpha_{\text{crit}}$ should be rejected or reinterpreted, and physically motivated priors on the shape parameters should be incorporated into fitting pipelines to avoid unphysical solutions. Enforcing the constraints on the core--S\'ersic model parameters is essential for obtaining reliable measurements of core structure, for interpreting these measurements in the context of galaxy formation and SMBH evolution, and for ensuring that the core--S\'ersic model continues to serve as a robust bridge between observations and theory.

\section{Conclusions}
\label{sec:conclusions}

The core--S\'ersic model has become the standard description of the surface-brightness profiles of massive early-type galaxies with partially depleted stellar cores. Although it is usually applied purely as a two-dimensional fitting function, its interpretation in terms of core sizes, mass deficits, and the evolution of SMBH binaries requires the underlying three-dimensional structure to correspond to a physically plausible stellar system. In this paper we have examined this issue in detail by analysing the intrinsic density profiles associated with the general core--S\'ersic family. Our main conclusions are as follows:
\begin{enumerate}
\item Numerical deprojections show that sharp transitions between the inner power-law core and outer S\'ersic component generically produce non-monotonic intrinsic density profiles. Such behaviour is unphysical for collisionless stellar systems and rules out the commonly adopted sharp-transition limit $\alpha\!\rightarrow\!\infty$.
\item For each combination of structural parameters $(\gamma, m, \Re/\Rb)$, there exists a critical transition parameter, $\alpha_{\text{crit}}$, above which the intrinsic density becomes non-monotonic. This value depends in a systematic way on the parameters: steeper S\'ersic indices and flatter inner slopes require particularly smooth transitions.
\item A substantial fraction of the formal core--S\'ersic parameter space is therefore physically inadmissible. Fits with $\alpha > \alpha_{\text{crit}}$ cannot correspond to realistic stellar systems, irrespective of the quality of the observational data.
\item These constraints have important consequences for measuring core sizes and mass deficits in massive ellipticals, for constructing dynamical models, and for comparing observations with simulations of SMBH binary scouring. Fits should employ physically motivated priors that restrict $\alpha$ to the admissible region of parameter space.
\end{enumerate}
By placing the core--S\'ersic model on a firmer physical footing, our results provide a more reliable basis for interpreting the central structure of massive galaxies and for connecting observed cores to the merger-driven evolution of their SMBHs.

\bibliography{mybib_nameyear}

\onecolumn
\appendix
\section{Numerical values for the critical transition parameter}
\label{appA.sec}

Table~{\ref{tab:alphacrit}} provides numerical values for the critical transition parameter $\alpha_{\text{crit}}$ for various combinations of the inner logarithmic slope $\gamma$, the ratio $\Re/\Rb$, and the S\'ersic index $m$. 

\begin{table*}[bh!]
\caption{Critical transition parameter $\alpha_{\text{crit}}$ as a function of model parameters $\gamma$, $\Re/\Rb$, and $m$.}
\label{tab:alphacrit}
\centering
\small
\setlength{\tabcolsep}{4pt}
\begin{tabular}{ccccccccccc}
\hline\hline\\[-0.7ex]
$\gamma$ & $\Re/\Rb$ & $m=2$ & $m=3$ & $m=4$ & $m=5$ & $m=6$ & $m=7$ & $m=8$ & $m=9$ & $m=10$ \\[1.5ex]
\hline
\\[-0.7ex]
0     &  $\ldots$    &  2.00    & 2.00    & 2.00    & 2.00  & 2.00    & 2.00    & 2.00    & 2.00    & 2.00    \\
0.025 &   5   & 3.67 & 3.55 & 3.49 & 3.46 & 3.44 & 3.43 & 3.42 & 3.41 & 3.41 \\
$\ldots$ &  10   & 3.94 & 3.68 & 3.58 & 3.53 & 3.50 & 3.47 & 3.46 & 3.44 & 3.43 \\
$\ldots$ &  15   & 4.13 & 3.78 & 3.64 & 3.57 & 3.53 & 3.50 & 3.48 & 3.46 & 3.45 \\
$\ldots$ &  20   & 4.30 & 3.85 & 3.69 & 3.60 & 3.55 & 3.52 & 3.50 & 3.48 & 3.46 \\
$\ldots$ &  30   & 4.58 & 3.96 & 3.76 & 3.65 & 3.59 & 3.55 & 3.52 & 3.50 & 3.48 \\
$\ldots$ &  50   & 5.03 & 4.13 & 3.85 & 3.72 & 3.64 & 3.59 & 3.55 & 3.53 & 3.51 \\
$\ldots$ &  70   & 5.40 & 4.26 & 3.92 & 3.76 & 3.67 & 3.62 & 3.58 & 3.55 & 3.52 \\
$\ldots$ & 100   & 5.90 & 4.41 & 4.00 & 3.82 & 3.71 & 3.65 & 3.60 & 3.57 & 3.54 \\
0.05  &   5   & 4.65 & 4.40 & 4.29 & 4.24 & 4.20 & 4.18 & 4.16 & 4.14 & 4.13 \\
$\ldots$  &  10   & 5.16 & 4.65 & 4.46 & 4.36 & 4.29 & 4.25 & 4.22 & 4.20 & 4.18 \\
$\ldots$  &  15   & 5.57 & 4.83 & 4.57 & 4.43 & 4.35 & 4.30 & 4.26 & 4.23 & 4.21 \\
$\ldots$  &  20   & 5.92 & 4.97 & 4.65 & 4.49 & 4.40 & 4.34 & 4.29 & 4.26 & 4.24 \\
$\ldots$  &  30   & 6.55 & 5.20 & 4.78 & 4.58 & 4.47 & 4.39 & 4.34 & 4.30 & 4.27 \\
$\ldots$  &  50   & 7.66 & 5.54 & 4.97 & 4.70 & 4.56 & 4.46 & 4.40 & 4.35 & 4.31 \\
$\ldots$  &  70   & 8.66 & 5.80 & 5.10 & 4.79 & 4.62 & 4.51 & 4.44 & 4.38 & 4.34 \\
$\ldots$  & 100   & 10.12& 6.13 & 5.26 & 4.90 & 4.69 & 4.57 & 4.48 & 4.42 & 4.37 \\
0.10  &   5   & 6.59 & 6.01 & 5.78 & 5.66 & 5.58 & 5.53 & 5.49 & 5.46 & 5.44 \\
$\ldots$  &  10   & 7.83 & 6.56 & 6.12 & 5.90 & 5.77 & 5.68 & 5.62 & 5.57 & 5.54 \\
$\ldots$  &  15   & 8.91 & 6.96 & 6.36 & 6.06 & 5.89 & 5.78 & 5.70 & 5.64 & 5.60 \\
$\ldots$  &  20   & 9.95 & 7.30 & 6.54 & 6.19 & 5.99 & 5.86 & 5.76 & 5.70 & 5.64 \\
$\ldots$  &  30   & 11.97& 7.85 & 6.83 & 6.38 & 6.13 & 5.97 & 5.85 & 5.77 & 5.71 \\
$\ldots$  &  50   & 16.27& 8.74 & 7.26 & 6.65 & 6.32 & 6.12 & 5.98 & 5.88 & 5.80 \\
$\ldots$  &  70   & 21.28& 9.47 & 7.59 & 6.85 & 6.47 & 6.23 & 6.06 & 5.95 & 5.86 \\
$\ldots$  & 100   & 31.02&10.41 & 7.98 & 7.09 & 6.63 & 6.35 & 6.16 & 6.03 & 5.93 \\
0.15  &   5   & 8.80 & 7.73 & 7.32 & 7.11 & 6.98 & 6.89 & 6.82 & 6.77 & 6.74 \\
$\ldots$ &  10   &11.25 & 8.70 & 7.90 & 7.52 & 7.29 & 7.14 & 7.04 & 6.96 & 6.90 \\
$\ldots$ &  15   &13.64 & 9.45 & 8.32 & 7.80 & 7.50 & 7.31 & 7.17 & 7.07 & 7.00 \\
$\ldots$ &  20   &16.15 &10.09 & 8.65 & 8.01 & 7.66 & 7.43 & 7.28 & 7.16 & 7.07 \\
$\ldots$ &  30   &21.87 &11.21 & 9.19 & 8.35 & 7.90 & 7.62 & 7.43 & 7.29 & 7.19 \\
$\ldots$ &  50   &38.21 &13.09 &10.00 & 8.84 & 8.24 & 7.88 & 7.64 & 7.46 & 7.33 \\
$\ldots$ &  70   &50.00 &14.77 &10.64 & 9.21 & 8.49 & 8.07 & 7.79 & 7.59 & 7.44 \\
$\ldots$ & 100   & ---  &17.11 &11.43 & 9.64 & 8.78 & 8.28 & 7.95 & 7.72 & 7.55 \\
0.20  &   5   &11.48 & 9.66 & 9.01 & 8.67 & 8.47 & 8.33 & 8.24 & 8.16 & 8.11 \\
$\ldots$ &  10   &15.97 &11.25 & 9.92 & 9.31 & 8.95 & 8.72 & 8.56 & 8.44 & 8.35 \\
$\ldots$ &  15   &21.00 &12.53 &10.59 & 9.74 & 9.27 & 8.97 & 8.77 & 8.61 & 8.50 \\
$\ldots$ &  20   &27.03 &13.68 &11.14 &10.09 & 9.52 & 9.17 & 8.92 & 8.75 & 8.61 \\
$\ldots$ &  30   &44.19 &15.75 &12.04 &10.64 & 9.90 & 9.46 & 9.16 & 8.94 & 8.78 \\
$\ldots$ &  50   & ---  &19.55 &13.46 &11.44 &10.45 & 9.87 & 9.48 & 9.21 & 9.01 \\
$\ldots$ &  70   & ---  &23.26 &14.62 &12.06 &10.86 &10.16 & 9.71 & 9.40 & 9.16 \\
$\ldots$ & 100   & ---  &28.99 &16.11 &12.81 &11.33 &10.50 & 9.97 & 9.61 & 9.34 \\
0.25  &   5   &14.83 &11.90 &10.91 &10.41 &10.11 & 9.91 & 9.77 & 9.67 & 9.58 \\
$\ldots$ &  10   &22.89 &14.37 &12.27 &11.34 &10.81 &10.47 &10.23 &10.06 & 9.93 \\
$\ldots$ &  15   &33.60 &16.49 &13.30 &11.99 &11.28 &10.84 &10.53 &10.31 &10.14 \\
$\ldots$ &  20   &49.18 &18.45 &14.16 &12.51 &11.65 &11.12 &10.76 &10.50 &10.31 \\
$\ldots$ &  30   & ---  &22.24 &15.63 &13.36 &12.23 &11.55 &11.10 &10.78 &10.55 \\
$\ldots$ &  50   & ---  &29.91 &18.02 &14.63 &13.06 &12.16 &11.58 &11.17 &10.88 \\
$\ldots$ &  70   & ---  &38.42 &20.06 &15.64 &13.69 &12.61 &11.92 &11.45 &11.11 \\
$\ldots$ & 100   & ---  & ---  &22.82 &16.87 &14.44 &13.13 &12.32 &11.77 &11.37 \\[1.5ex]
\hline
\end{tabular}
\tablefoot{For each combination of these parameters, models with $\alpha > \alpha_{\text{crit}}$ have non-monotonic intrinsic density profiles and are therefore physically inadmissible. Missing entries (---) indicate that $\alpha_{\text{crit}}$ lies outside the explored range (i.e. larger than 50).}
\end{table*}

\addtocounter{table}{-1}
\begin{table*}
\caption{(continued)}
\centering
\small
\setlength{\tabcolsep}{4pt}
\begin{tabular}{ccccccccccc}
\hline\hline\\[-0.7ex]
$\gamma$ & $\Re/\Rb$ & $m=2$ & $m=3$ & $m=4$ & $m=5$ & $m=6$ & $m=7$ & $m=8$ & $m=9$ & $m=10$ \\[1.5ex]
\hline
\\[-0.7ex]
0.30  &   5   &19.13 &14.53 &13.07 &12.36 &11.94 &11.66 &11.47 &11.32 &11.21 \\
$\ldots$ &  10   &33.67 &18.32 &15.06 &13.68 &12.92 &12.43 &12.10 &11.86 &11.67 \\
$\ldots$ &  15   & ---  &21.74 &16.61 &14.63 &13.59 &12.95 &12.52 &12.21 &11.97 \\
$\ldots$ &  20   & ---  &25.10 &17.93 &15.40 &14.12 &13.35 &12.84 &12.47 &12.20 \\
$\ldots$ &  30   & ---  &32.07 &20.25 &16.67 &14.96 &13.97 &13.33 &12.87 &12.54 \\
$\ldots$ &  50   & ---  &48.28 &24.27 &18.64 &16.21 &14.86 &14.01 &13.43 &13.00 \\
$\ldots$ &  70   & ---  & ---  &27.86 &20.23 &17.16 &15.53 &14.51 &13.83 &13.33 \\
$\ldots$ & 100   & ---  & ---  &33.06 &22.27 &18.32 &16.30 &15.09 &14.28 &13.70 \\
0.35  &   5   &24.84 &17.69 &15.58 &14.58 &14.00 &13.62 &13.35 &13.15 &13.00 \\
$\ldots$ &  10   & ---  &23.39 &18.42 &16.42 &15.34 &14.67 &14.21 &13.88 &13.62 \\
$\ldots$ &  15   & ---  &28.97 &20.71 &17.78 &16.28 &15.38 &14.78 &14.35 &14.03 \\
$\ldots$ &  20   & ---  &34.80 &22.76 &18.91 &17.04 &15.94 &15.23 &14.72 &14.34 \\
$\ldots$ &  30   & ---  &48.08 &26.42 &20.78 &18.25 &16.81 &15.90 &15.27 &14.80 \\
$\ldots$ &  50   & ---  & ---  &33.16 &23.80 &20.08 &18.10 &16.87 &16.04 &15.44 \\
$\ldots$ &  70   & ---  & ---  &39.70 &26.34 &21.52 &19.06 &17.58 &16.60 &15.90 \\
$\ldots$ & 100   & ---  & ---  &49.71 &29.70 &23.29 &20.22 &18.41 &17.25 &16.42 \\
0.40  &   5   &32.69 &21.52 &18.51 &17.13 &16.33 &15.82 &15.46 &15.19 &14.99 \\
$\ldots$ &  10   & ---  &30.16 &22.54 &19.66 &18.16 &17.23 &16.61 &16.16 &15.82 \\
$\ldots$ &  15   & ---  &39.36 &25.92 &21.59 &19.46 &18.21 &17.38 &16.80 &16.36 \\
$\ldots$ &  20   & ---  &49.74 &29.03 &23.22 &20.53 &18.98 &17.98 &17.29 &16.78 \\
$\ldots$ &  30   & ---  & ---  &34.92 &25.99 &22.24 &20.20 &18.92 &18.04 &17.41 \\
$\ldots$ &  50   & ---  & ---  &46.58 &30.66 &24.92 &22.02 &20.27 &19.10 &18.28 \\
$\ldots$ &  70   & ---  & ---  & ---  &34.73 &27.07 &23.42 &21.27 &19.88 &18.90 \\
$\ldots$ & 100   & ---  & ---  & ---  &40.33 &29.80 &25.11 &22.46 &20.79 &19.63 \\
0.45  &   5   &43.81 &26.24 &21.99 &20.08 &19.01 &18.32 &17.85 &17.50 &17.23 \\
$\ldots$ &  10   & ---  &39.43 &27.71 &23.57 &21.46 &20.20 &19.36 &18.77 &18.32 \\
$\ldots$ &  15   & ---  & ---  &32.68 &26.26 &23.25 &21.52 &20.40 &19.62 &19.03 \\
$\ldots$ &  20   & ---  & ---  &37.47 &28.61 &24.73 &22.58 &21.21 &20.27 &19.59 \\
$\ldots$ &  30   & ---  & ---  &47.17 &32.73 &27.19 &24.28 &22.49 &21.29 &20.42 \\
$\ldots$ &  50   & ---  & ---  & ---  &40.04 &31.13 &26.83 &24.36 &22.74 &21.60 \\
$\ldots$ &  70   & ---  & ---  & ---  &46.66 &34.37 &28.85 &25.80 &23.84 &22.47 \\
$\ldots$ & 100   & ---  & ---  & ---  & ---  &38.58 &31.37 &27.48 &25.08 &23.47 \\
0.50  &   5   & ---  &32.18 &26.13 &23.54 &22.10 &21.20 &20.56 &20.11 &19.76 \\
$\ldots$ &  10   & ---  & ---  &34.22 &28.30 &25.39 &23.68 &22.57 &21.76 &21.18 \\
$\ldots$ &  15   & ---  & ---  &41.71 &32.10 &27.85 &25.46 &23.95 &22.89 &22.11 \\
$\ldots$ &  20   & ---  & ---  &49.29 &35.54 &29.90 &26.90 &25.03 &23.76 &22.84 \\
$\ldots$ &  30   & ---  & ---  & ---  &41.73 &33.40 &29.25 &26.77 &25.12 &23.97 \\
$\ldots$ &  50   & ---  & ---  & ---  & ---  &39.18 &32.88 &29.35 &27.10 &25.55 \\
$\ldots$ &  70   & ---  & ---  & ---  & ---  &44.10 &35.85 &31.40 &28.62 &26.72 \\
$\ldots$ & 100   & ---  & ---  & ---  & ---  & ---  &39.55 &33.79 &30.37 &28.11 \\[1.5ex]
\hline
\end{tabular}
\end{table*}

\end{document}